\documentclass[prl,aps,twocolumn,10pt]{revtex4-1}
\usepackage[utf8x]{inputenc}
\usepackage{amsmath,amssymb,graphicx,xcolor,esint,bm}

\newcommand{\Esp}{{\bf E}_{S}^{+}}
\newcommand{\Esm}{{\bf E}_{S}^{-}}
\newcommand{\Ei}{{\bf E}_{i}}

\newcommand{\Mi}[3]{{\bf M}_{#3} ^{\left(#1\right)} \left( #2 \right)}
\newcommand{\Ni}[3]{{\bf N}_{#3}^{\left(#1\right)} \left( #2 \right)}
\newcommand{\Mii}[2]{{\bf M}_{#2}^{\left(#1\right)}}
\newcommand{\Nii}[2]{{\bf N}_{#2}^{\left(#1\right)}}

\newcommand{\Sqrta}{\sqrt{\alpha_{nl}}}
\newcommand{\re}[1]{\mbox{Re}\left\{#1\right\}}
\newcommand{\no}{\hat{\bf n}}
\newcommand{\im}[1]{\mbox{Im}\left\{#1\right\}}
\newcommand{\Sqrtb}{\sqrt{\beta_{nl}}}
\newcommand{\J}[2]{j_{#2}\left(#1\right)}
\newcommand{\h}[2]{h_{#2}^{\left( 1 \right)}\left(#1\right)}

\newif\ifpaper
\paperfalse

\begin{document}

\title{Material-independent modes for electromagnetic scattering}

\author{Carlo Forestiere}
\affiliation{ Department of Electrical Engineering and Information Technology, Universit\`{a} degli Studi di Napoli Federico II, via Claudio 21,
 Napoli, 80125, Italy}
\author{Giovanni Miano}
\affiliation{ Department of Electrical Engineering and Information Technology, Universit\`{a} degli Studi di Napoli Federico II, via Claudio 21,
 Napoli, 80125, Italy}
%\section{Formulation}
%\
\begin{abstract}
In this Letter, we introduce a representation of the electromagnetic field for the analysis and synthesis of the full-wave scattering by a homogeneous dielectric object of arbitrary shape in terms of a set of eigenmodes independent of its permittivity. The expansion coefficients are rational functions of the permittivity.  This approach naturally highlights the role of plasmonic and photonic modes in any scattering process and suggests a straightforward methodology to design the permittivity of the object to pursue a  prescribed tailoring of the scattered field. 
We discuss in depth the application of the proposed approach to the analysis and design of the scattering properties of a dielectric sphere.
 
\end{abstract}
\maketitle
The advent of metamaterials drove a fundamental paradigm change in the design of electromagnetic devices: permittivity and permeability are no longer confined to the range of values dictated  by the materials found in nature, but rose to the role of true design parameters  in a spectral range  spanning from microwaves to optical frequencies \cite{Pendry06,Leonhardt06,silva2014,jahani2016all,Khorasaninejad1190}. 
Nevertheless, the formulations of the Maxwell's equations currently used for the analysis and design of electromagnetic scattering processes did not adjust to this paradigm shift and have remained substantially unaltered for decades, with the exceptions of transformation optics \cite{Pendry06,Leonhardt06}. In fact, in most of the  approaches used for solving the Maxwell's equations the  contributions of the material parameters  and of the geometry are mathematically intertwined and cannot be separated. Therefore, the design of a material to achieve assigned constraints on the scattered electromagnetic field is very complicated. 
For instance, in the Mie theory, which provides an analytical description of the interaction of the electromagnetic field with a sphere \cite{Mie08,stratton07}, the permittivity and the sphere's radius both appear in the argument of the vector spherical wave functions, and, as a consequence, the Mie expansion coefficients are complicated functions of their combination. Only recently, has this problem  been addressed in the case of { scalar} Mie scattering  by Markel \cite{Markel10}.

In this Rapid Communication, by using spectral methods, we derive an approach for the analysis and synthesis of the electromagnetic scattering from a dielectric object of arbitrary shape that gives prominence to the material properties, by distinctly separating the role they have in the scattering processes from the role played by the geometry. In particular,  we reduce the vector scattering problem to an
algebraic form  by introducing an auxiliary eigenvalue problem.

Our approach
naturally leads to that developed in Refs. \cite{Bergman78,Fredkin2003,Mayergoyz05} in the
quasi-static limit, and to that proposed by Markel \cite{Markel10} in the case of scalar Mie scattering.
An analogous approach has been introduced in Ref. \cite{Bergman80}, even though it was only applied  to isolated and interacting spheres in the limit of small particle radii compared to the wavelength. Very recently it has also been used in a one-dimensional case to describe the full-wave electromagnetic response of a flat-slab composite structure \cite{Bergman16}.
 
Then, we demonstrate that the design of metamaterials can be greatly simplified by expressing the scattered electromagnetic field in
terms of material-independent eigenmodes, exploiting the fact that the expansion coefficients are a rational
function of the permittivity. %
As an example, we analytically investigate the spectral properties of the electromagnetic scattering from a sphere, showing the universal loci described by the eigenvalues as a function of the sphere's  size parameter, systematizing within this framework the properties of plasmonic and photonic modes.
 Then, we design the permittivity of the sphere to have vanishing backscattering. In the case of an arbitrarily shaped object, the evaluation of the eigenvalues and the corresponding eigenmodes can be carried out by using standard numerical tools.

Let us consider the electromagnetic scattering by an object occupying a regular region $\Omega$ with boundary $\partial \Omega$. The object is excited by a time harmonic electromagnetic field incoming from infinity $\re{\Ei \left({\bf r}\right) e^{- i \omega t}}$. The medium is a non-magnetic isotropic homogeneous dielectric with relative permittivity $\varepsilon_R \left( \omega \right)$, surrounded by vacuum. Let $\Esp$ and $\Esm$ be the scattered electric fields in $\Omega$ and $\mathbb{R}^3 \backslash \bar{\Omega}$, respectively. The Maxwell’s equations lead to
\begin{alignat}{2}
\label{eq:MErotp}
& \boldsymbol{\nabla}^2 \Esp + k_0^2 \varepsilon_R \left( \omega \right) \Esp = k_0^2 \left[ 1 - \varepsilon_R \left( \omega \right) \right] \Ei  \; && \mbox{in} \, \Omega, \\
\label{eq:MErote}
& \boldsymbol{\nabla}^2 \Esm + k_0^2  \Esm = {\bf 0}  \; && \mbox{in} \,\mathbb{R}^3 \backslash \bar{\Omega}, \\
\label{eq:BC}
   & \no \times \left( \Esm - \Esp \right) = {\bf 0} \; && \mbox{on} \, \partial \Omega, \\
   \label{eq:BCrot}
   & \no \times \left( \boldsymbol{\nabla} \times \Esm - \boldsymbol{\nabla} \times\Esp \right) = {\bf 0} \; && \mbox{on} \, \partial \Omega,   
\end{alignat}
where $k_0= \omega/c_0$, $c_0$ is the light velocity in vacuum and $\no$ is the outgoing normal to $\partial \Omega$.  Equations \ref{eq:MErotp}-\ref{eq:BCrot} have to be solved with the  radiation conditions, namely the regularity and Silver-M\"uller conditions at infinity \cite{SM}. This problem has a unique solution if $\im {\varepsilon_R} >0 $ \cite{cessenat1996}.

{ Aiming at the reduction of the scattering problem to an algebraic form}, we introduce the following auxiliary eigenvalue problem
\begin{alignat}{2}
\label{eq:CalderonNabla}
 & -{k_0^{-2}} \, \boldsymbol{\nabla}^2 {\bf C}  =  \gamma {\bf C} \qquad &&  \mbox{in} \; \Omega, \\
 \label{eq:Calderon}
& \hat{\bf n} \times \boldsymbol{\nabla} \times {\bf C}  = \mathcal{C}^{e} \left\{ \hat{\bf n} \times {\bf C} \right\} \qquad && \mbox{in} \; \partial \Omega,
\end{alignat}
where $\gamma$ is the eigenvalue. We introduced the exterior outgoing Calder\'on operator $\mathcal{C}^{e}$ \cite{cessenat1996} that takes the tangential component   of the scattered electric field on $\partial \Omega$, i.e. $\left. \hat{\bf n} \times \Esm \right|_{\partial \Omega} $, where $\Esm$ is solution of the scattering problem, and returns the tangential component of its curl $ \left. \hat{\bf n} \times \boldsymbol{\nabla} \times \Esm \right|_{\partial \Omega} $:
\begin{equation}
  \mathcal{C}^{e} \left\{  \left. \hat{\bf n} \times \Esm  \right|_{\partial \Omega} \right\} = \left. \no \times \boldsymbol{\nabla} \times \Esm \right|_{\partial \Omega}.
\end{equation}
The Calder\'on operator only depends on the geometry of $\partial \Omega$. 
Since the operator $-\boldsymbol{\nabla}^2$ in $\Omega$ with the boundary condition \ref{eq:Calderon} is compact, its spectrum $\left\{ \gamma_r \right\}_{r \in \mathbb{N}}$ is countably infinite. 
 This fact is a consequence of the radiation conditions, which are implicitly accounted for by the exterior Calder\'on operator.
 
  In this case, the operator $-\boldsymbol{\nabla}^2$ is not Hermitian (even though symmetric), thus its eigenvalues are complex with $\im {\gamma_r} < 0 $. The  eigenmodes ${\bf C}_r$ and ${\bf C}_s$ corresponding to different eigenvalues $\gamma_r$ and $\gamma_s$ are not orthogonal in the usual sense, i.e. $\langle {\bf C}_r^*,{\bf C}_s\rangle_\Omega \ne 0$, where
 \begin{equation}
\langle \mathbf{A},\mathbf{B} \rangle_V = \iiint_V \mathbf{A} \cdot \mathbf{B} \, \mbox{dV}.
\end{equation}
 Nevertheless, by introducing its dual operator \cite{SM} it can be proved that
\begin{equation}
 \langle {\bf C}_r,{\bf C}_s\rangle_\Omega = 0 \qquad  \gamma_{r} \ne \gamma_{s},
 \label{eq:Orthogonality}
\end{equation}
and
\begin{equation}
 \gamma_r = \frac{1}{\left\| {\bf C}_r \right\|^2_{\Omega}} 
 \left[\frac{\left\| \boldsymbol{\nabla} \times {\bf C}_r \right\|^2_{\mathbb{R}^3}}{k_0^2} -\left\| {\bf C}_r \right\|^2_{\mathbb{R}^3 \backslash \bar{\Omega}}
 - i
 \varoiint_{S_\infty} \hspace{-1em}  \frac{\left| {\bf C}_r \right|^2}{k_0}
 \mbox{dS}
    \right]
 \label{eq:Balance}
\end{equation}
where $\left\| {\bf A} \right\|_{V}^2 = \langle {\bf A}^*, {\bf A} \rangle_{V}$. The eigenmodes ${\bf C}_r$ are extended in $\mathbb{R}^3$ by requiring that they satisfy Eq. \ref{eq:MErote}, the boundary conditions \ref{eq:BC}-\ref{eq:BCrot} and the radiation conditions at infinity. 
Although the property \ref{eq:Orthogonality} is shared with the so called quasi normal modes (QNMs) \cite{Ching1998}, they are  fundamentally different \cite{SM}. Contrarily to the introduced modes, the QNMs depend on the permittivity of the object and diverge at infinity \cite{Kristensen13}.

 Equation \ref{eq:Balance} suggests that $\re{\gamma_r}$ does not have a definite sign, while $\im{\gamma_r}$ is strictly negative. In particular, $\im{\gamma_r}$ is proportional to the contribution of the corresponding mode to the power radiated to infinity, accounting for its radiative losses.

In the presence of an arbitrary external excitation $\Ei$, the solution of the scattering problem is
\begin{equation}
   \Esp = \left( 1 - \varepsilon_R \right) \displaystyle \sum_{r=1}^\infty \frac{1}{\varepsilon_R - \gamma_r} \frac{\langle {\bf C}_r, {\bf E}_i \rangle_\Omega}{\langle {\bf C}_r, {\bf C}_r \rangle_\Omega} {\bf C}_r
   \label{eq:Exp}
\end{equation}
The eigenvalues $\gamma_r$ and the eigenfunctions ${\bf C}_r$ are  permittivity independent, and they only depend on the geometry of the dielectric object. The permittivity appears in the multiplicative factors only as $ 1/\left( \varepsilon_R - \gamma_r \right)$.

From now on, we assume that the region $\Omega$ is a sphere of radius $R$, with   size parameter  $x= 2 \pi \, {R}/{\lambda}$, where $\lambda$ is the  wavelength. The set of eigenvalues $\left\{ \gamma_r \right\}_{r \in \mathbb{N}}$ is the union of  $\left\{ \alpha_{nl} \right\}_{ \left( n,l \right) \in \mathbb{N}^2}$ and $\left\{ \beta_{nl} \right\}_{ \left( n,l \right) \in \mathbb{N}^2}$ being $\alpha_{nl}$ (respectively $\beta_{nl}$) the $l$-th root of the power series $\mathcal{P}_n$ (respectively $\mathcal{Q}_n$) \cite{SM}:
\begin{equation}
   \mathcal{P}_n \left( \alpha \right) = \sum_{h=0}^{\infty} \;  p_{nh} \, \left( \alpha -1 \right)^h,  \quad
   \mathcal{Q}_n \left( \beta  \right) =\sum_{h=0}^{\infty} \;  q_{nh} \, \left( \beta -1 \right)^h,
   \label{eq:Poly}
\end{equation}
where the coefficients $p_{nh}\left( x \right)$ and $q_{nh}\left( x \right)$ are defined for any given $n$, $h \ge 1 $, and $x$ as follows:
\begin{equation*}
\begin{aligned}
&p_{n0} = \, q_{n0} = \h{x}{n+1} \J{x}{n} - \h{x}{n} \J{x}{n+1}, \\
&q_{nh}= - \frac{\left(- 1 \right)^{h-1}}{\left(h-1\right)!}   \left(\frac{x}{2}\right)^{h-1} \left[ \h{x}{n} \J{x}{n+h}  \right] + \\ &
  \frac{\left(- 1 \right)^h}{h!}   \left(\frac{x}{2}\right)^h  \left[ \h{x}{n+1} \J{x}{n+h} - \h{x}{n} \J{x}{n+h+1}  \right], \\
&p_{nh}  = q_{nh} - \frac{\left(- 1 \right)^{h-1}}{\left(h-1\right)!} \left(\frac{x}{2}\right)^{h-1} \frac{ \left[ x \h{x}{n} \right]'}{x} \J{x}{n+h-1},	
\end{aligned}
\end{equation*}
where $j_n$ and $h_n$ are the spherical Bessel and Hankel functions of the first kind, respectively. The eigenspace corresponding to the eigenvalue $\alpha_{nl}$ is spanned by the eigenmodes $\Ni{1}{\Sqrta k_0 {\bf r}}{ \substack{e \\ o} mn}$ with $m \in \mathbb{N}_0$, which feature zero radial magnetic field. Therefore, they are denoted as {\it electric type} modes. Dual reasoning leads us to call the eigenmodes $\Mi{1}{\Sqrtb k_0 {\bf r}}{ \substack{e \\ o} mn}$ associated with the eigenvalues $\beta_{nl}$ {\it magnetic type} modes. The radial mode number $l$ gives the number of maxima along $\hat{\bf r}$ inside the sphere.  The functions $\Nii{1}{ \substack{e \\ o} mn}$  and $\Mii{1}{ \substack{e \\ o} mn}$
are the vector spherical wave functions (VSWFs) regular at the origin \cite{SM} and the subscripts $e$ and $o$ denote even and odd azimuthal dependence. 

The solution of the electromagnetic scattering problem is:
\begin{equation}
\begin{aligned}
   &{\bf E}_S^+ \left( {\bf r} \right) = \left( \varepsilon_R - 1 \right)
\displaystyle\sum_{mnl} \left( \frac{B_{emnl}}{ \beta_{nl} - \varepsilon_R } \Mi{1}{\Sqrtb \, k_0  {\bf r}}{emn} + \right. \\ & \frac{B_{omnl}}{ \beta_{nl} - \varepsilon_R } \Mi{1}{\Sqrtb \, k_0 {\bf r}}{omn} +
 \frac{A_{emnl}}{ \alpha_{nl} - \varepsilon_R } \Ni{1}{ \Sqrta \, k_0 {\bf r}}{emn} + \\  & \left. \frac{A_{omnl}}{ \alpha_{nl} - \varepsilon_R } \Ni{1}{ \Sqrta \, k_0 {\bf r}}{omn} \right),
\end{aligned}
\label{eq:ExpansionEi}
\end{equation}
where $\displaystyle\sum_{nml} =
 \displaystyle\sum_{n=1}^\infty \displaystyle\sum_{m=0}^n  \displaystyle\sum_{l=1}^\infty \,$, and 
\begin{equation}
\begin{aligned}
A_{ \substack{e \\ o} m n l} &= \frac{\langle 
\Ni{1}{\Sqrta \, k_0  {\bf r}}{ \substack{e \\ o} mn}, {\bf E}_i \rangle_\Omega}{\langle 
\Ni{1}{\Sqrta \, k_0  {\bf r}}{ \substack{e \\ o} mn}, \Ni{1}{\Sqrta \, k_0 {\bf r}}{ \substack{e \\ o} mn} \rangle_\Omega}, \\
B_{ \substack{e \\ o} m n l} &= \frac{\langle 
\Mi{1}{\Sqrtb \, k_0  {\bf r}}{ \substack{e \\ o} mn}, {\bf E}_i \rangle_\Omega}{\langle 
\Mi{1}{\Sqrtb \, k_0  {\bf r}}{ \substack{e \\ o} mn}, \Mi{1}{\Sqrtb \, k_0 {\bf r}}{ \substack{e \\ o} mn}\rangle_\Omega}.
\end{aligned}
\label{eq:EiCoeff}
\end{equation}
 The coefficients introduced in Eq. \ref{eq:EiCoeff} feature an analytical expression in the case of an incident plane wave, reported in the supplemental material \cite{SM}.
 
  In passive materials where $\im{\varepsilon_R} \ge 0$, the quantities $\left| \alpha_{nl} - \varepsilon_R \right|$ and $\left| \beta_{nl} - \varepsilon_R \right|$ do not vanish as $\omega$ varies because $\im{\alpha_{nl}}<0$ and $\im{\beta_{nl}}<0$. Nevertheless, the mode amplitudes $ {A_{ \substack{e \\ o} m n l}}/{ \left( \alpha_{nl} - \varepsilon_R  \right) }$ and $ {B_{ \substack{e \\ o} m n l}}/{ \left( \beta_{nl} - \varepsilon_R \right)}$ reach their maximum whenever:
\begin{equation}
     \left| \varepsilon_R \left( \omega \right)  - \alpha_{nl}\left( x \right) \right| = \underset{x,\omega}{\mbox{min}}; \; 
     \left| \varepsilon_R \left( \omega \right) - \beta_{nl} \left( x \right) \right| = \underset{x,\omega}{\mbox{min}},
   \label{eq:Match}
\end{equation}
respectively.  These are the resonant conditions.

Since $\alpha_{nl}$ and $\beta_{nl}$ are independent of the sphere's permittivity, we exhaustively summarize their behavior by the loci they span in the  complex plane by varying the size parameter $x$. The resulting diagrams are universal, being valid for every conceivable homogeneous sphere. The loci belong to the half-plane $\mbox{Im} \left\{ \gamma_r \right\}<0$ (see Eq. \ref{eq:Balance}) because of the radiation condition at infinity. The real part of $\gamma_r$ can assume in general both positive and negative values. When it is negative the resonant condition \ref{eq:Match} is verified by metals at optical frequencies ($ \mbox{Re} \left\{ \varepsilon_R \right\} <0$), giving rise to plasmon oscillations (e.g. Ref. \cite{Mayergoyz05}). When it is positive the resonant condition $\ref{eq:Match}$ is verified by dielectrics ($ \mbox{Re} \left\{ \varepsilon_R \right\} \ge 0$), giving rise to photonic resonances.

\begin{figure}
\centering
\includegraphics[width=85mm]{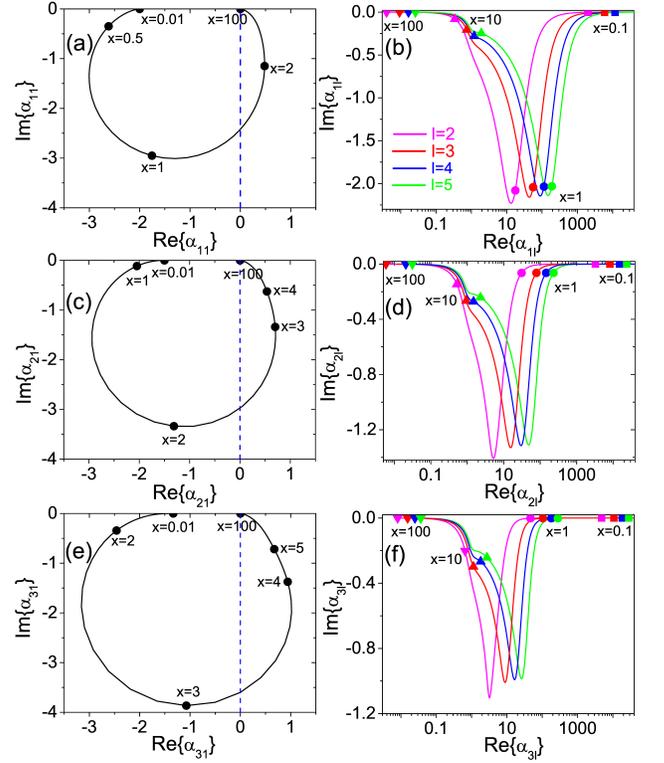}
\caption{Universal loci spanned in the  complex plane by the eigenvalues $\alpha_{nl}$ of the electric-type eigenmodes of a dielectric sphere by varying its   size parameter $x \in \left[0.01,100 \right]$. We show the  eigenvalues of the fundamental (a) and higher order  dipole modes (b), fundamental (c) and higher order (d) quadrupole modes, fundamental (e) and higher order (f) octupole modes. The panels (a,c,e) are in linear scale. The panels (b,d,f) are in semilog scale.}
  \label{fig:Figure_1}
\end{figure}

 In practice, we plot them by finding the roots of the two polynomials obtained by truncating the power series in Eq. \ref{eq:Poly} to $h_{max}=50$.
 First, we investigate the  eigenvalue $\alpha_{11}$.  The  spatial distribution of the corresponding eigenmodes $\Ni{1}{ \Sqrta k_0 {\bf r}}{\substack{e \\ o} m1}$ suggests the dipole character of this mode \cite{SM}.
 In Fig. \ref{fig:Figure_1} (a) we plot the locus spanned by $\alpha_{11}$. First, we note that for $x \ll 1$ the  eigenvalue $\alpha_{11}$ approaches the value $-2$, in accordance with the Fr\"{o}hlich condition \cite{kreibig2013}. This is consistent with Eq. \ref{eq:Balance} that shows that $\re{\gamma_n}<0$ in the quasi-electrostatic limit where $\boldsymbol{\nabla} \times {\bf C}_n \approx {\bf 0} $.
By increasing  $x$, both the real and the imaginary part of $\alpha_{11}$  move toward  more negative values. For Drude metals with low losses, this fact implies the  red shift of the corresponding resonant frequency \cite{maier07}. When $x\approx 0.72$ the quantity $\re{\alpha_{11}}$  reaches a minimum and then starts increasing.
For larger $x$, ${\alpha_{11}}$ lies in fourth quadrant of the  complex plane. Then,  $\re{\alpha_{11}}$  increases until $x\approx2$ where it reaches the maximum value of $0.48$, then $\alpha_{11}$ asymptotically approaches the origin of the  complex plane. Figure \ref{fig:Figure_1} (a) sets specific constraints on the permittivity that a homogeneous sphere should have to exhibit the fundamental dipole resonance. In particular, in the limit of low-losses its permittivity should satisfy the constraint $ -3 \le \re{\varepsilon_R} \le 0.48$.

The loci spanned by higher order electric dipole modes $\alpha_{1l}$ with $l=2,3,4,5$, shown in Fig. \ref{fig:Figure_1} (b),  manifest a very different nature. First, $\alpha_{1l}$ always lies in the fourth quadrant of the  complex plane irrespectively of the mode order $l \ge 2$.  Moreover, for $x \rightarrow 0$ the real part of $\alpha_{1l} \rightarrow \infty$, while $\mbox{Im} \left\{ \alpha_{1l} \right\}$ approaches zero. This fact means that for $x \ll 1$ these modes cannot be practically excited.  By increasing $x$, the  value $\mbox{Re} \left\{ \alpha_{1l} \right\}$ moves toward smaller values, while the imaginary part decreases and reaches a minimum. Eventually, $ \alpha_{1l}$ approaches the origin of the complex plane for very high values of $x$.

In Figs. \ref{fig:Figure_1} (c) and (e) we plot the loci spanned by $\alpha_{21}$ and $\alpha_{31}$ of the fundamental ($l=1$) electric quadrupole  and octupole   eigenmodes. In this case for $x \rightarrow 0$ the eigenvalues  $\alpha_{21}$ and $\alpha_{31}$ approach respectively the values $-1.5$ and $-1.33$, which agree with  the quasi-static approximation \cite{Fredkin2003,Mayergoyz05}. In Figs. \ref{fig:Figure_1} (d) and (f) we show the loci of higher order quadrupole and octupole modes which display a behavior similar to higher order dipole modes.
\begin{figure}
\centering
\includegraphics[width=85mm]{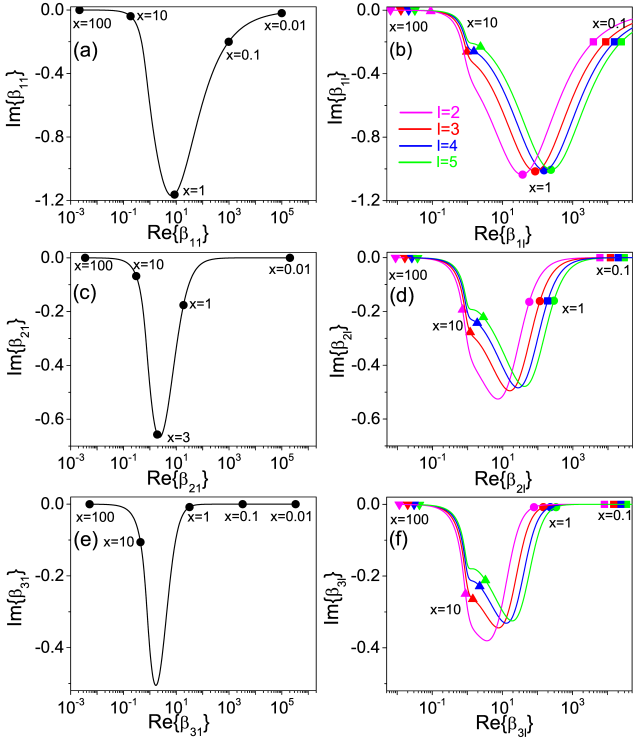}
\caption{Universal loci spanned in the  complex plane by the eigenvalues $\beta_{nl}$ of the magnetic-type eigenmodes of a dielectric sphere  by varying its size parameter $x \in \left[0.01,100 \right]$. We show the eigenvalues of the fundamental (a) and higher order (b) dipole modes, fundamental (c) and higher order (d) quadrupole modes, fundamental (e) and higher order (f) octupole modes. The panels (a,c,e) are in linear scale. The panels (b,d,f) are in semilog scale.}
  \label{fig:Figure_2}
\end{figure}

Let us now consider the  eigenvalues $\beta_{nl}$ of the magnetic-type eigenmodes. The  eigenvalues of both the fundamental magnetic-type eigenmodes, i.e. $\beta_{n1}$ shown in Fig. \ref{fig:Figure_2} (a,c,e) for $n=1,2,3$, and higher order magnetic eigenmodes, i.e.  $\beta_{nl}$ shown in Fig. \ref{fig:Figure_2} (b,d,f) for $l=2,3,4,5$, exhibit the same behavior of the  eigenvalue of higher order electric modes. In particular, in the limit for $x \rightarrow 0$ the quantity $\re{\beta_{n1}}$ diverges. Therefore, the fundamental magnetic eigenmodes cannot be practically excited in the electrostatic limit, consistently with Refs. \cite{Fredkin2003,Mayergoyz05}.

In conclusion, the only eigenmodes that can be resonantly excited in a metal sphere  with $\re{\varepsilon_R}<0$ regardless of $x$ are the fundamental electric ones. Moreover, only these modes have their loci confined in a limited region of the complex plane and are excitable in the electrostatic limit.  We thus identify them with the {\it plasmonic} modes, generalizing to the electrodynamic case the definition given in Ref. \cite{Fredkin2003}.  Conversely, we call the remaining eigenmodes {\it photonic} modes because they cannot be excited at low frequency.  The photonic modes are responsible for the ripple structure that can be observed in the extinction of large weakly absorbing spheres   \cite{bohren08}. 

\begin{figure}
\centering
\includegraphics[width=85mm]{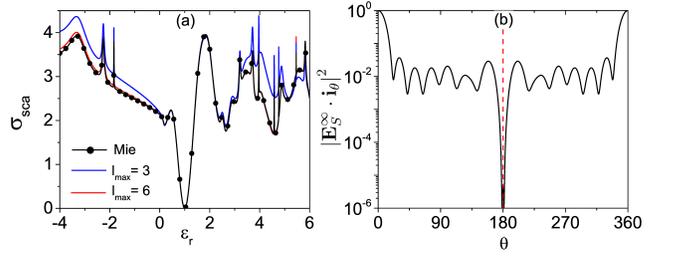}
\caption{(a) Scattering efficiency $\sigma_{sca}$ of a dielectric sphere with size parameter $x=2\pi$ excited by a linearly polarized plane wave, as a function of $\varepsilon_R \in \left[ -4, 6 \right]$ calculated using Eq. \ref{eq:ScattEff} assuming $l_{max}=3$ and $l_{max}=6$ and with the standard Mie theory. In all the calculations we have assumed $n_{max}=10$. (b) Squared magnitude of the radiation pattern for $\phi=0$ as a function of the angle $\theta$ for the sphere designed to have a vanishing back-scattering.}
  \label{fig:Figure_3}
\end{figure}

On the basis of the proposed modal expansion, we now  calculate the scattering efficiency $\sigma_{sca}$ of a sphere \cite{bohren08}, when it is excited by a linearly polarized plane wave:
\begin{multline}
   \sigma_{sca} = 2 \frac{\left( \varepsilon_R - 1 \right)^2 }{ x^2}   \sum_{n=1}^\infty  \frac{\left( 2 n +1 \right)}{\left({\h{x}{n}}\right)^2}  \times \\   
     \left[ \left|  \sum_{l=1}^{\infty} \frac{{ \Sqrta  \, \J{\Sqrta  x}{n}} A_{nl}}{ \alpha_{nl} - \varepsilon_R }  \right|^2 +  \left|  \sum_{l=1}^{\infty}   \frac{\J{\Sqrtb x}{n} B_{nl}}{ \beta_{nl} - \varepsilon_R } \right|^2 \right] 
   \label{eq:ScattEff}
\end{multline}
where:
\begin{equation}
\begin{aligned}
A_{n l} &= \frac{\langle 
\Ni{1}{\Sqrta k_0 {\bf r}}{e1n}, \Ni{1}{k_0 {\bf r}}{e1n} \rangle_\Omega}{\langle 
\Ni{1}{\Sqrta k_0 {\bf r}}{e1n}, \Ni{1}{\Sqrta k_0 {\bf r}}{e1n} \rangle_\Omega}, \\
B_{nl} &= \frac{\langle 
\Mi{1}{\Sqrtb k_0 {\bf r}}{o1n}, \Mi{1}{k_0 {\bf r}}{o1n}\rangle_\Omega}{\langle 
\Mi{1}{\Sqrtb k_0 {\bf r}}{ o1n}, \Mi{1}{\Sqrtb k_0 {\bf r}}{o1n}\rangle_\Omega}, 
\end{aligned}
\label{eq:CouplingPW}
\end{equation}
(see the supplemental material for the corresponding analytical expressions). Specifically, in Fig. \ref{fig:Figure_3} (a)  we  plot $\sigma_{sca}$ for $x=2\pi$ as a function of $\varepsilon_R$, calculated by truncating  the exterior sum of Eq. \ref{eq:ScattEff} to $n_{max}=10$, and the inner sum to $l_{max}=3$ (blue line) and to $l_{max}=6$ (red line). We compare them with the standard Mie solution calculated assuming the same value of $n_{max}$. Although for $l_{max}=3$ it is apparent that there is a moderate disagreement with the Mie theory, for $l_{max}=6$ the outcomes of the two approaches become almost indistinguishable. Differently from the Mie theory, our method allows one to directly associate each peak to the modes that have generated it.

Now, let us put our material-independent modal decomposition to practical use. Suppose we want to cancel the backscattering of a homogeneous sphere of a given  size parameter by designing its permittivity. We assume that the sphere has radius $R=\lambda$, i.e. $x=2\pi$, and is excited by a $x$-polarized plane wave of unit intensity, propagating along the $z$-axis. The answer to this question is straightforward by using the developed approach and only requires one to find the roots of a polynomial equation. The radiation pattern is defined as
$
 \mathbf{E}_S^\infty \left( \theta, \phi, \varepsilon_R \right) = \displaystyle\lim_{r \rightarrow \infty}  \left[ r{e^{- i k_0 r}} \Esm \right]
$, where $\theta$ and $\phi$ are the polar and azimuthal angles, respectively.
Due to symmetry considerations the only non-vanishing component of the radiation pattern in the backscattering direction ($\theta = \pi$) is $\mathbf{E}_S^\infty \cdot {\bf i}_\theta$. Therefore, our task is to find the zeros of $\mathbf{E}_S^\infty \cdot {\bf i}_\theta$ as a function of $\varepsilon_R$, where $\mathbf{E}_S^\infty \cdot {\bf i}_\theta$ is expressed as \cite{SM}:
\begin{equation}
\begin{aligned}
   {\bf E}_S^\infty \cdot {\bf i}_\theta & = \frac{\varepsilon_R - 1 }{k_0} \displaystyle\sum_{n=1}^{\infty} \sum_{l=1}^{\infty}   \left\{ \frac{\delta_{nl}\left( x, \theta, \phi \right)}{ \beta_{nl} \left( x \right) - \varepsilon_R }  +  \frac{\gamma_{nl}\left( x, \theta, \phi \right)}{ \alpha_{nl}\left( x \right) - \varepsilon_R } \right\}
\end{aligned}
\label{eq:diffSca}
\end{equation}
where 
\begin{equation*}
\begin{aligned}
   \gamma_{nl} \left( x, \theta, \phi \right) &=  -i E_n \frac{ \Sqrta  \, \J{\Sqrta  x}{n}}{{\h{x}{n}}} A_{nl}  
\Ni{\infty}{\theta,\phi}{e1n}  \cdot {\bf i}_\theta,  \\
   \delta_{nl} \left( x, \theta, \phi \right) &=  E_n  \frac{ \J{\Sqrtb x}{n}}{{\h{x}{n}}}   B_{nl} \Mi{\infty}{\theta,\phi}{o1n} \cdot {\bf i}_\theta, 
\end{aligned}
\end{equation*}
$E_n = i^n \left( 2n + 1\right) / \left[ n \left( n + 1 \right) \right]$, $A_{nl}$ and $B_{nl}$ are defined in Eq. \ref{eq:CouplingPW}, $
 \Mii{\infty}{o1n} = \displaystyle\lim_{  r \rightarrow \infty}  \left[ k_0 {r} {e^{- i k_0 r}} \Mii{3}{o1n} \right]
$,
$
 \Nii{\infty}{e1n} = \displaystyle\lim_{r \rightarrow \infty}  \left[ k_0 {r} {e^{- i k_0 r}} \Nii{3}{e1n} \right]
$, and the functions $\Nii{3}{ \substack{e \\ o} mn}$  and $\Mii{3}{ \substack{e \\ o} mn}$ are the  radiative VSWFs.
We set $x=2\pi$, $\theta=\pi$ and $\phi=0$ in the expression \ref{eq:diffSca} truncated with $n_{max}=10$ and $l_{max}=8$. Then, we put all the terms in the sum of Eq. \ref{eq:BCrot} over a common denominator, obtaining in this way a rational function  and we zero the resulting numerator, which is a polynomial in $\varepsilon_R$. Among the different solutions, we chose the only one that is physically realizable by a passive material, i.e. $\varepsilon_R = -2.2746969 + 0.0818799 i $. To validate this result, we plot in Fig. \ref{fig:Figure_3} (b) the squared magnitude of the radiation pattern, i.e. the differential scattering cross section, of the designed sphere as a function of the angle $\theta$ for $\phi=0$ computed by using the standard Mie theory with $n_{max}=10$. 
We achieved a ratio between the back- and the forward- scattered power of -53dB. It is worth noting that the achieved backscattering suppression cannot be attributed to known interference conditions such as the Kerker conditions \cite{Kerker83,Person13}, but originates from a complex interplay of many electric and magnetic scattering orders, which are significant up to $n=9$.

  This method can be also used to design the  permittivity of the sphere to pursue many different goals, including zeroing or focusing a given field component in an arbitrary point of space inside or outside the volume $\Omega$, in the near or in the far zone. These objectives can be all easily achieved by zeroing a polynomial.  Finally, we note that the proposed method leads to a high computational burden when $x \gg 1$ because many modes have to be considered to accurately describe the field.

\begin{acknowledgements}
We thank the anonymous reviewers for bringing Refs. \cite{Markel10,Bergman16} to our attention.
\end{acknowledgements}

\end{document}